# Computational aberration compensation by coded aperture-based correction of aberration obtained from Fourier ptychography (CACAO-FP)

Jaebum Chung[1,*], Gloria W. Martinez[2], Karen Lencioni[2], Srinivas Sadda[3], and Changhuei Yang[1]

[1]Department of Electrical Engineering, California Institute of Technology, Pasadena, CA 91125, USA
[2]Office of Laboratory Animal Resources, California Institute of Technology, Pasadena, CA 91125, USA
[3]Doheny Eye Institute, University of California – Los Angeles, Los Angeles, CA 90033, USA

Abstract

    We report a novel generalized optical measurement system and computational approach to determine and correct aberrations in optical systems. We developed a computational imaging method capable of reconstructing an optical system's aberration by harnessing Fourier ptychography (FP) with no spatial coherence requirement. It can then recover the high resolution image latent in the aberrated image via deconvolution. Deconvolution is made robust to noise by using coded apertures to capture images. We term this method: coded aperture-based correction of aberration obtained from Fourier ptychography (CACAO-FP). It is well-suited for various imaging scenarios with the presence of aberration where providing a spatially coherent illumination is very challenging or impossible. We report the demonstration of CACAO-FP with a variety of samples including an in-vivo imaging experiment on the eye of a rhesus macaque to correct for its inherent aberration in the rendered retinal images. CACAO-FP ultimately allows for a poorly designed imaging lens to achieve diffraction-limited performance over a wide field of view by converting optical design complexity to computational algorithms in post-processing.

Introduction

    A perfect aberration-free optical lens simply does not exist in reality. As such, all optical imaging systems constructed from a finite number of optical surfaces are going to experience some level of aberration issues. This simple fact underpins the extraordinary amount of optical design efforts that have gone into the design of optical imaging systems. In broad terms, optical imaging system design can be defined as the complex process by which specialized optical elements and their spatial relationships are chosen in order to minimize aberrations and provide an acceptable image resolution over a desired field of view (FOV) [1]. The more optical surfaces available to the designer, the greater the extent the aberrations can be minimized. However, this physical system improvement approach for minimizing aberrations has reached a point of diminishing returns in modern optics. Microscope objectives with 15 optical elements have become commercially available in recent years [2], but it is difficult to imagine that it would be feasible to cram another order of magnitude more optical surfaces into the confines of an objective in the foreseeable future. Moreover, this strategy for minimizing aberration is/can never be expected to accomplish the task of completely zeroing out aberrations. In other words, any optical system's spatial bandwidth product, which scales as the product of system FOV and inverse resolution, can be expected to remain a design bound dictated by the residual aberrations in the system.

    The issue of aberrations in simpler optical systems with few optical surfaces is, unsurprisingly, more pronounced. The eye is a very good example of such an optical system. While it does a fair job of conveying external scenes onto our retinal layer, its optical quality is actually quite poor. When a clinician desires a high resolution of the retinal layer itself for diagnostic purposes, the human eye lens and cornea aberrations would have to be somehow corrected or compensated for. The prevalent approach by which this is currently done is through the use of adaptive optics (AO) [3, 4]. This is in effect a sophisticated way of physically correcting aberrations where complex physical optical elements are used to compensate for the aberrations of the lens and cornea. Although AO can be fast and accurate, its performance is limited to the small region around the aberration-compensated spot within the lens's FOV. Since a lens's aberration is spatially variant, AO needs to be raster-scanned to obtain an aberration-free image across the lens's FOV [5]. In addition, AO system requires an active feedback loop with a wavefront measurement device and a compensation device, which can dramatically complicate the optical system [6].

Fourier ptychography (FP) circumvents the challenges associated with physically correcting aberrations by recasting the problem of aberration correction as a computational problem that can be solved after image data has been acquired. Rather than striving to get the highest quality images possible through an imaging system, FP acquires a controlled set of aberrated images, dynamically determines the system's aberration characteristics computationally, and reconstitutes a high-quality aberration-corrected image from the original controlled image set [7-13]. One way to view FP is to note its similarity to synthetic aperture imaging. In a standard FP microscope system, images of the target are collected through a low numerical aperture (NA) objective with the target illuminated with a series of angularly varied planar or quasi-planar illumination. Viewed in the spatial frequency domain, each image represents a disc of information with its offset from the origin determined by the illumination angle. As with synthetic aperture synthesizing, we then stitch the data from the collected series in the spatial frequency domain. Unlike synthetic aperture imaging, we do not have direct knowledge of the phase relationships between each image data set. In FP, we employ phase retrieval and the partial information overlap amongst the image set to converge on the correct phase relationships during the stitching process [7]. At the end of the process, the constituted information in the spatial frequency domain can be Fourier transformed to generate a higher resolution image of the target that retains the original FOV as set by the objective. It has been demonstrated that a sub-routine can be weaved into the primary FP algorithm that will dynamically determine the aberration pupil function of the imaging system [9]. In fact, the majority of existing FP algorithms incorporate some versions of this aberration determination function to find and subsequently correct out the aberrations from the processed image [14-19]. This particular sub-discipline of FP has matured to the level that it is even possible to use a very crude lens to obtain high quality images that are typically associated with sophisticated imaging systems [20] – this drives home the fact that correcting aberration computationally is a viable alternative to physical correction.

The primary objective of this paper is to report a novel generalized optical measurement system and computational approach to determine and correct aberrations in optical systems. This computational approach is coupled to a general optical scheme designed to efficiently collect the type of image required by the computational approach. Currently, FP's ability to determine and correct aberration is limited to optical setups with well-defined, spatially coherent field on the sample plane [7, 9, 22-30]. We developed a computational imaging method capable of reconstructing an optical system's aberration by harnessing FP with no spatial coherence requirement. It can then recover the high resolution image latent in the aberrated image via deconvolution. Deconvolution is made robust to noise by using coded apertures to capture images [31]. We term this method: coded aperture-based correction of aberration obtained from Fourier ptychography (CACAO-FP). It is well-suited for various imaging scenarios with the presence of aberration where providing a spatially coherent illumination is very challenging or impossible. CACAO-FP ultimately allows for a poorly designed imaging lens to achieve diffraction-limited performance over a wide FOV by converting optical design complexity to computational algorithms in post-processing.

The removal of spatial coherence constraints is vitally important in allowing us to apply computation aberration correction to a broader number of imaging scenarios. These scenarios include: 1) optical systems where the illumination on a sample is provided via a medium with unknown index variations; 2) optical systems where space is so confined that it is not feasible to employ optical propagation to create quasi-planar optical fields; 3) optical systems where the optical field at the sample plane is spatially incoherent by nature (e.g. fluorescence emission).

CACAO-FP is substantially different from other recent efforts aimed at aberration compensation. Broadly speaking, these efforts can be divided into two major categories: blind and heuristic aberration recovery. Blind recovery minimizes a cost function, typically an image sharpness metric or a maximum-likelihood function, over a search space, usually the coefficient space of Zernike orthonormal basis [32-37], to arrive at the optimal aberration function. However, blind recovery is prone to converging towards multiple local minima, and requires the aberrated sample image to be a complex field because blind aberration recovery with intensity-only sample image is extremely prone to noise for any aberration [33] other than a linear blur [38]. Heuristic recovery algorithms rely on several assumptions, such as assuming that the captured complex-field sample image has diffuse distribution in its Fourier spectrum such that each sub-region in the Fourier domain encodes the local aberrated wavefront information [39-42]. Thus, heuristic methods are limited to specific types of samples and their performance is highly sample dependent.

CACAO-FP is capable of achieving a robust aberration recovery performance in a generalized and broadly applicable format. In the first section, we describe the principle of CACAO-FP. In section 2, we report the demonstration of CACAO-FP with a crude lens and an eye phantom as imaging systems of interest. Finally, in section 3, we report the use of CACAO-FP in an in-vivo imaging experiment on the eye of a rhesus macaque to correct for its inherent aberration in the rendered retinal images.

Section 1

To best understand the overall operation of the CACAO-FP processing, we start by examining the optical scheme (see Fig. 1). Suppose we start with an unknown optical system of interest (target system). This target system consists of a lens (unknown lens) placed at focal length in front of a target sample (unknown sample). The sample is illuminated incoherently. For simplicity in this thought experiment, we will consider the illumination to occur in the transmission mode. The CACAO-FP system collects light from the target system using relay lenses $L_{s1}$, $L_{s2}$, and $L_{s3}$, and an aperture mask in the Fourier plane, with its coordinates $(u, v)$, that can be modulated into different patterns. Our objective is to resolve the sample at high resolution. It should be clear from this target system description that our ability to achieve the objective is confounded by the presence of the unknown lens and its unknown aberrations. A good example of such a system is the eye - the retinal layer is the unknown sample, and the lens and cornea can be represented by the unknown lens.

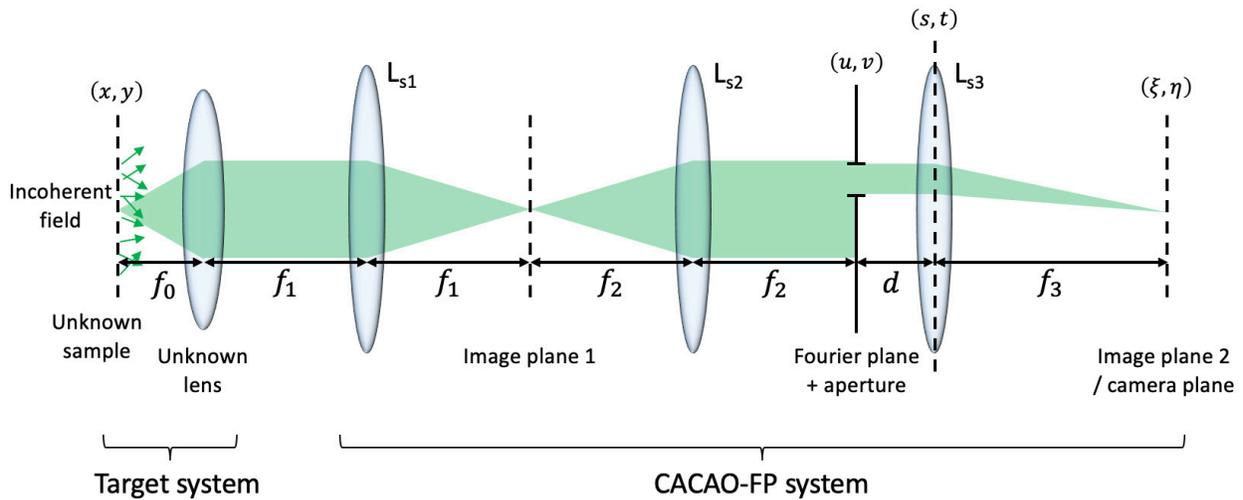

Figure 1. Optical architecture of CACAO-FP. CACAO-FP system consists of 3 tube lenses to relay the image from the target system for analysis. The target system consists of an unknown lens and an unknown sample with spatially incoherent field. CACAO-FP system has access to the Fourier plane that can be arbitrarily modulated with binary patterns using a spatial light modulator. The images captured by CACAO-FP system is intensity-only.

From this thought experiment, we can see that, to accomplish our objective, we would need to first determine the aberration characteristics of the unknown lens and then use the information to somehow correct out the aberration effects from the final rendered image. CACAO-FP does this by using 3 primary computational imaging algorithmic components that operate in sequence: 1) local aberration recovery with blur estimation; 2) full aberration recovery with FP; and 3) latent image recovery by deconvolution with coded apertures. The first two steps determine the target system's aberrations, and the third step generates an aberration-corrected image. This pipeline is summarized in Fig. 2. The sample plane, with its coordinates $(x, y)$, is divided into small tiles within which the aberration can be assumed to be spatially invariant, and CACAO-FP processes each corresponding tile on its image plane, with coordinates $(\xi, \eta)$, to recover a high resolution image of the sample tile. In the following analysis, we focus our attention to one tile, $t0$. CACAO-FP begins by capturing a series of images with varying mask patterns in its Fourier plane, with its coordinates $(u, v)$. The patterns consist of two kinds: a set of small circular apertures, $W_m(u, v)$, that spans the pupil of the unknown lens; and a set of big apertures, $A_n(u, v)$, that includes coded apertures and a full circular aperture with their diameters equal to the unknown lens's pupil size. $m$ and $n$ are integers from 1 to the total number of the respective aperture. The images captured with $W_m(u,v)$ are labeled as $i_{m,t0}(\xi,\eta)$, and they encode the local aberration of the unknown lens's pupil function in their PSFs. The blur estimation algorithm (Section 1.1) extracts these PSFs, $b_{m,t0}(\xi,\eta)$. These intensity values of the spatially filtered pupil function can be synthesized with Fourier ptychography (Section 1.2) into the full pupil function, $P_{t0}(u,v)$. The images captured with $A_n(u,v)$, labeled $\phi_{n,t0}(\xi,\eta)$, are processed with the reconstructed pupil function and the knowledge of the mask patterns to generate the latent, aberration-free image of the sample, $o_{t0}(x,y)$ (Section 1.3).

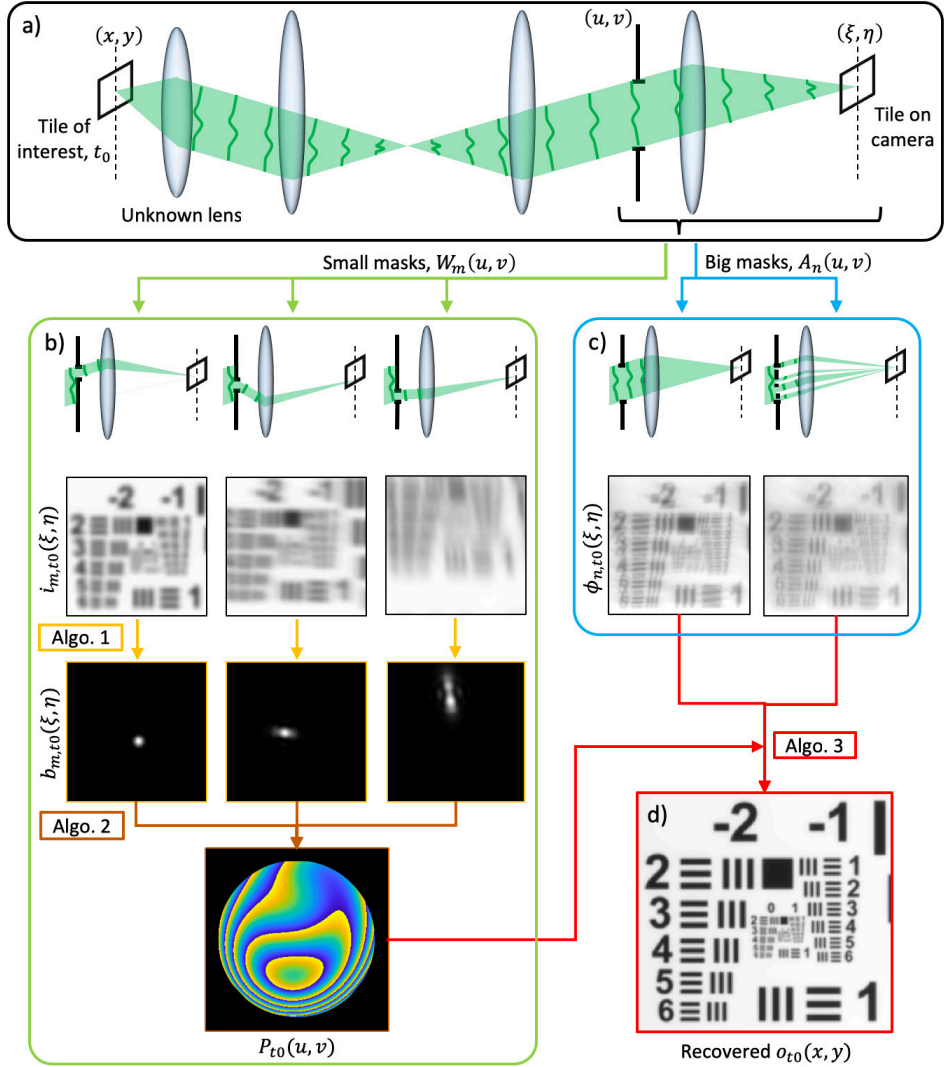

Figure 2. Outline of CACAO-FP pipeline. a) The captured images are broken into small tiles of isoplanatic patches (i.e. aberration is spatially invariant within each tile). b) Data acquisition and post-processing for estimating the pupil function, $P_{t0}(u,v)$. Limited-aperture images, $i_{m,t0}(\xi,\eta)$, are captured with small masks, $W_m(u,v)$, applied at the Fourier plane. Local PSFs, $b_{m,t0}(\xi,\eta)$, are determined by the blur estimation procedure, Algorithm 1. These PSFs are synthesized into the full-aperture pupil function, $P_{t0}(u,v)$, with Fourier ptychography, Algorithm 2. c) Data acquisition with big masks, $A_n(u,v)$, at the Fourier plane. The recovered $P_{t0}(u,v)$ from b) and the big-aperture images $\phi_{n,t0}(\xi,\eta)$ from c) are used for deconvolution, Algorithm 3, to recover the latent aberration-free intensity distribution of the sample, $o_{t0}(x,y)$.

The next 3 sub-sections will explain the 3 imaging algorithmic components and the associated data collection process in detail.

Section 1.1 – Local aberration recovery with blur estimation

We consider a point on the unknown sample, $s(x,y)$, and how it propagates to the camera plane to be imaged. On the sample plane, a point source at $(x_0, y_0)$ may have an amplitude and phase $C$, and it can be described by:

$$U_0(x, y; x_0, y_0) = C\delta(x - x_0, y - y_0)$$
Eq. 1

We then use Fresnel propagation and apply the phase delay associated with an idealized thin lens having an estimated focal length $f_0$ for the unknown lens [Appendix 1.1]. Any discrepancy from the ideal is incorporated into the pupil function, $P(u, v; x_0, y_0)$, which is usually a circular bandpass filter with a uniform modulus and a phase modulation. Thus, the field right after passing through the unknown lens is:

$$U_F(u, v; x_0, y_0) = C_F(x_0, y_0)P(u, v; x_0, y_0)\exp\left[-j\frac{2\pi}{\lambda f_0}(x_0 u + y_0 v)\right]$$
Eq. 2

where $C_F(x_0, y_0)$ is a constant factor, $\lambda$ is the wavelength of the field, and $(u, v)$ are the coordinates of both the plane right after the unknown lens and the Fourier plane as these planes are conjugate to each other. The spatially varying nature of a lens's aberration is captured by the pupil function's dependence on $(x_0, y_0)$. We divide our sample into small tiles (e.g. $t0, t1, t2, \ldots$) and confine our analysis to one tiled region, $t0$, on the sample plane that contains $(x_0, y_0)$ and other points in its vicinity such that the spatially varying aberration can be assumed to be constant, $P(u, v; x_0, y_0) = P_{t0}(u, v)$, from here on. This is a common strategy for processing spatially variant aberration in a wide FOV imaging [43, 44]. We can see from Eq. 2 that the field emerging from the unknown lens is essentially its pupil function with the phase gradient defined by the point source's location on the sample plane.

At the Fourier plane, a user-defined aperture mask, $M(u, v)$, is applied to produce:

$$U'_F(u, v; x_0, y_0) = M(u, v)P_{t0}(u, v)\exp\left[-j\frac{2\pi}{\lambda f_0}(x_0 u + y_0 v)\right]$$
Eq. 3

where we dropped the constant factor $C_F(x_0, y_0)$. After further propagation to the camera plane [Appendix 1.1], we obtain the intensity pattern, $i_{PSF,t0}(\xi, \eta)$, that describes the mapping of a point on the sample to the camera plane:

$$i_{PSF,t0}(\xi, \eta; x_0, y_0) = \left|\mathcal{F}\{M(u, v)P_{t0}(u, v)\}(\xi, \eta) * \delta\left(\xi + \frac{x_0}{\lambda f_0}, \eta + \frac{y_0}{\lambda f_0}\right)\right|^2 = h_{t0}\left(\xi + \frac{x_0}{\lambda f_0}, \eta + \frac{y_0}{\lambda f_0}\right)$$
Eq. 4

where $h_{t0}(\xi, \eta) = |\mathcal{F}\{M(u, v)P_{t0}(u, v)\}(\xi, \eta)|^2$ is the intensity of the point-spread-function (PSF) of the combined system in Fig. 1 for a given aperture mask $M(u, v)$ and within the isoplanatic patch $t0$. We observe that PSFs for different point source locations are related to each other by simple lateral shifts, such that the image, $i_{t0}(\xi, \eta)$, of an unknown sample within the isoplanatic patch, $s_{t0}(x, y)$, captured by this system can be represented by:

$$i_{t0}(\xi, \eta) = h_{t0}(\xi, \eta) * |s_{t0}(\xi, \eta)|^2 = h_{t0}(\xi, \eta) * o_{t0}(\xi, \eta)$$
Eq. 5

where $o_{t0}(\xi, \eta)$ is the intensity of $s_{t0}(\xi, \eta)$, and we ignore the coordinate scaling for simplicity. This equation describes that the image captured by the detector is a convolution of the sample's intensity field with a PSF associated with the sub-region of the pupil function defined by an arbitrary mask at the pupil plane. This insight allows us to capture images of the sample under the influence of PSFs that originate from different sub-regions of the pupil. We have aperture masks of varying shapes and sizes. To avoid confusion, we label the mth small mask, its associated PSF in isoplanatic patch $t0$, and the image captured with it in the local PSF determination procedure as $W_m(u, v)$, $b_{m,t0}(\xi, \eta)$, and $i_{m,t0}(\xi, \eta)$, respectively, and the nth big mask (coded aperture or a full aperture), its associated PSF, and image as $A_n(u, v)$, $h_{n,t0}(\xi, \eta)$, and $\phi_{n,t0}(\xi, \eta)$, respectively.

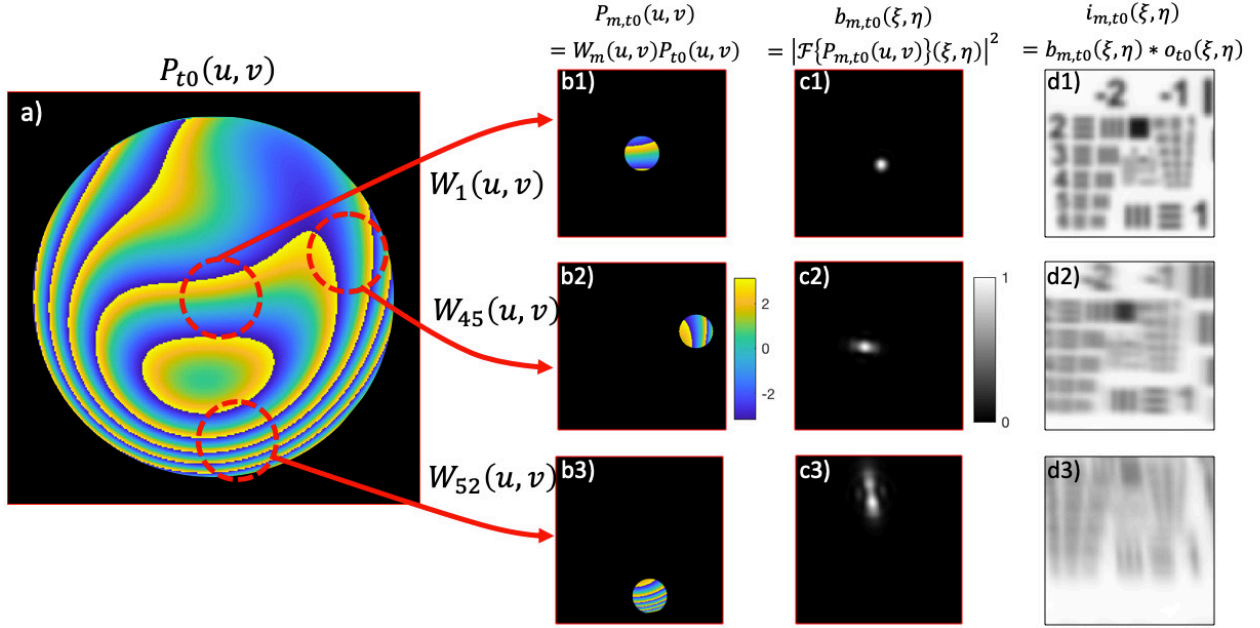

Figure 3. Simulating image acquisition with different small masks at the Fourier plane. a) The full pupil function masked by the lens's NA-limited aperture. Differently masked regions of the pupil, b1-3), give rise to different blur kernels, c1-3), which allows us to capture images of the sample under the influence of different PSFs. Only the phase is plotted for $P_{t0}(u,v)$ and $P_{m,t0}(u,v)$'s, and their apertures are marked by the black boundaries. $W_1(u,v)$, $W_{45}(u,v)$, and $W_{52}(u,v)$ are 3 small masks from a spiraling-out scanning sequence.

As shown in Fig. 3, a local PSF, $b_{m,t0}(\xi,\eta)$, has a spatial translation according to the general phase gradient direction of the masked region, and its shape can deviate from a point with more severe aberration within the masked region. In general, the aberration at or near the center of an imaging lens is minimal, and it becomes severe near the edge of the aperture because the lens's design poorly approximates the parabolic shape away from the optical axis [45]. Thus, the image captured with the center mask, $i_{1,t0}(\xi,\eta)$, is mostly aberration-free with its PSF defined by the diffraction-limited spot associated with the mask's aperture size. Other $i_{m,t0}(\xi,\eta)$'s are under the influence of additional aberration encapsulated by its local PSF, $b_{m,t0}(\xi,\eta)$.

Our CACAO-FP pipeline begins with data acquisition required to determine the pupil function, $P_{t0}(u,v)$, as described in Fig. 2 b). We capture a series of $i_{m,t0}(\xi,\eta)$ with small masks, $W_m(u,v)$, at the Fourier plane. Here, we adopt an image-pair-based blur estimation algorithm widely used in computational photography discipline to determine $b_{m,t0}(\xi,\eta)$. In this algorithm, one of the image pair is assumed to be blur-free while the other is blurred [46, 47]. The blur kernel can be estimated by an iterative PSF estimation method, which is iterative Tikhonov deconvolution [48] in Fourier domain, adopting update scheme in Yuan's blur estimation algorithm [46] and adjusting the step size to be proportional to $\frac{|I_{1,t0}(u,v)|}{|I_{1,t0}(u,v)|_{\max}}$ for robustness to noise [49], where $I_{1,t0}(u,v)$ is the Fourier spectrum of $i_{1,t0}(\xi,\eta)$. The blur estimation process is described in Algorithm 1.

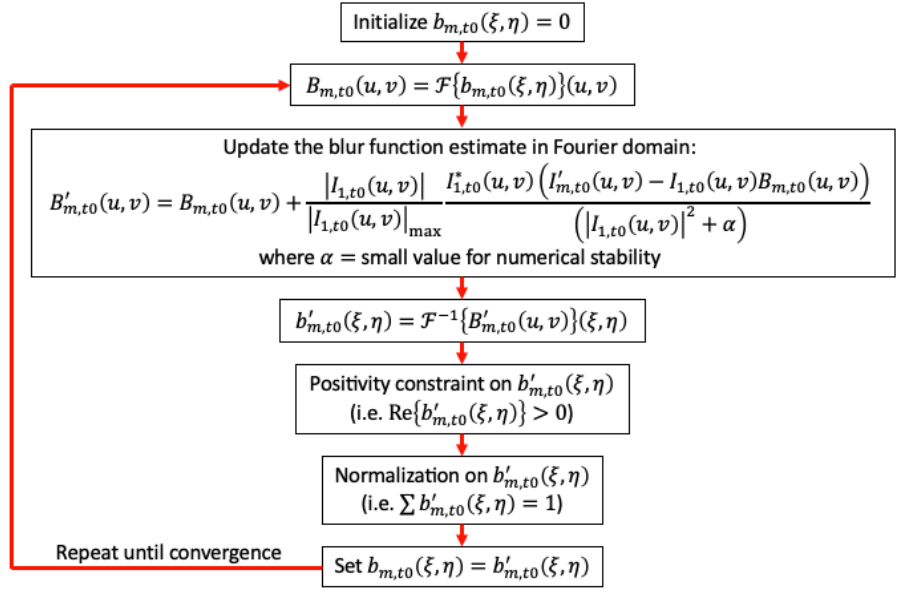

Algorithm 1. Blur estimation algorithm for determining local PSFs from images captured with small apertures, $W_{m,t0}(u,v)$.

The recovered local PSFs, $b_{m,t0}(\xi,\eta)$, are the intensity information of the different masked pupil regions' Fourier transform. They can be synthesized into the full pupil function, $P_{t0}(u,v)$, using Fourier ptychography, as described in the following section.

Section 1.2 – Full aberration recovery with Fourier ptychography (FP)

FP uses a phase retrieval algorithm to synthesize a sample's Fourier spectrum from a series of intensity images of the sample captured by scanning an aperture on its Fourier spectrum [7, 8]. In our implementation of FP, the full pupil's complex field, $P_{t0}(u,v)$, is the desired Fourier spectrum to be synthesized, and the local PSFs, $b_{m,t0}(\xi,\eta)$ are the aperture-scanned intensity images to be used for FP, as shown in the bottom half of Fig. 2 b). The FP algorithm is described in Algorithm 2.

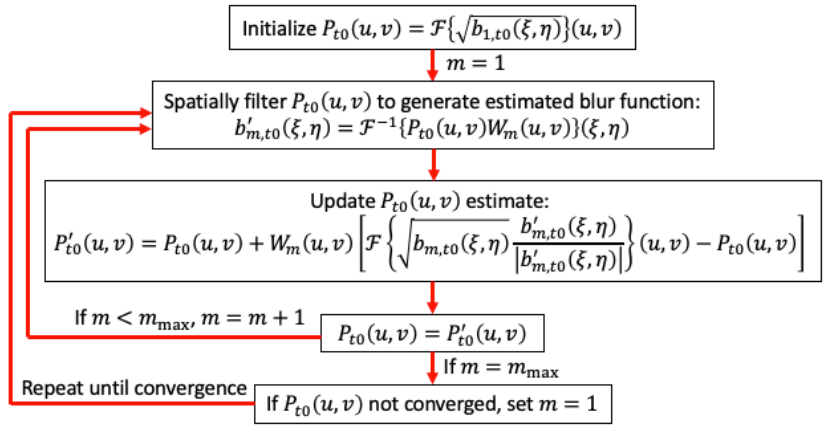

Algorithm 2. Fourier ptychography algorithm for reconstructing the unknown lens's pupil function, $P_{t0}(u,v)$.

Fourier ptychography requires that the scanned apertures during image acquisition have at least 30%

overlap [50] for successful phase retrieval. Thus, the updating $b_{m,t0}(\xi,\eta)$ in Algorithm 2 are ordered in a spiral-out pattern, each having an associated aperture $W_m(u,v)$ that partially overlaps (40% by area) with the previous one's aperture.

Section 1.3 – Latent image recovery by deconvolution with coded apertures

With the knowledge of the pupil function obtained from FPM, it is possible to recover $o_{t0}(x,y)$ from the aberrated image $\phi_{t0}(\xi,\eta)$ taken with the full pupil aperture. In the Fourier domain, the image's spectrum is represented as: $\Phi_{t0}(u,v) = H_{t0}(u,v)O_{t0}(u,v)$, where $H_{t0}(u,v)$ and $O_{t0}(u,v)$ are the spatial spectrum of $h_{t0}(\xi,\eta)$ and $o_{t0}(x,y)$, respectively. $H_{t0}(u,v)$ is also called the optical transfer function (OTF) of the optical system and, by Fourier relation, is an auto-correlation of the pupil function $P_{t0}(u,v)$. In the presence of severe aberrations, the OTF may have values at or close to zero for many spatial frequency regions within the bandpass, as shown in Fig. 4. These are due to the phase gradients with opposite slopes found in an aberrated pupil function, which may produce values at or close to zero in the auto-correlation process. Thus, the division of $\Phi_{t0}(u,v)$ by $H_{t0}(u,v)$ during deconvolution will amplify noise at these spatial frequency regions since the information there has been lost in the image acquisition process. This is an ill-posed inverse problem.

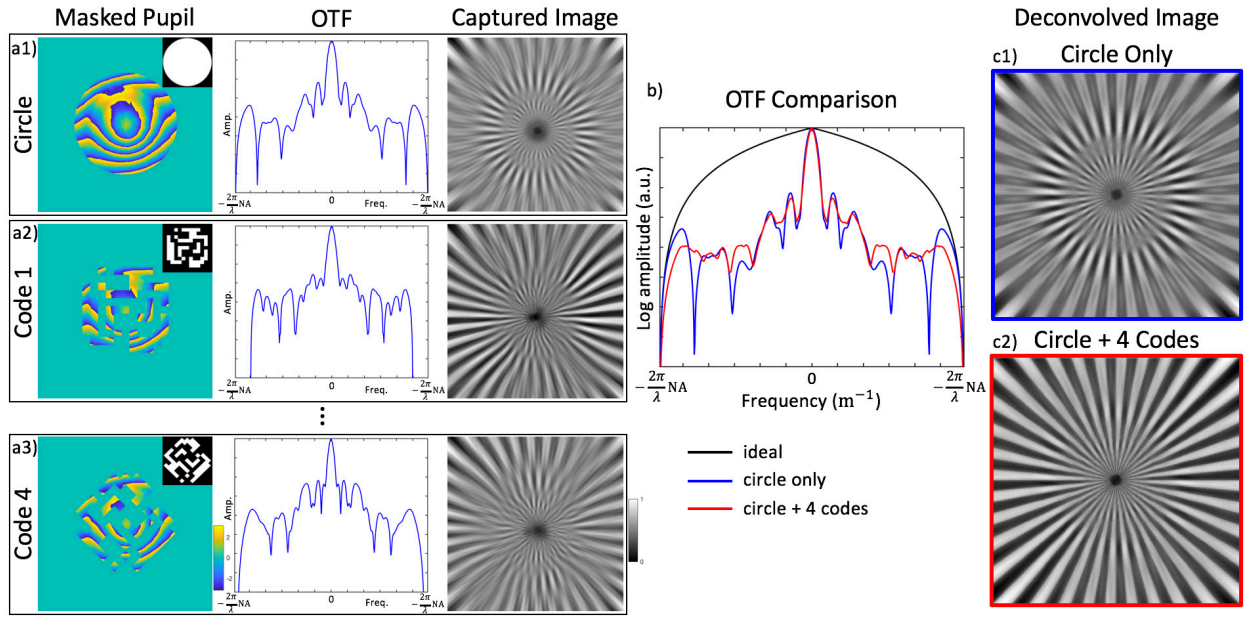

Figure 4. Simulation that demonstrates the benefit of coded aperture-based deconvolution. a1-3) shows the pupil function masked by the full circular aperture and coded apertures under different rotation angles, their associated OTFs and captured images. b) compares the OTF of the ideal pupil function with the OTF of a full pupil function and the summed OTFs of the pupil masked by a full aperture and 4 coded apertures. Null regions in the frequency spectrum are avoided by using coded apertures, allowing for better deconvolution results as shown in c2) compared to a full aperture-only deconvolution c1).

There are several deconvolution methods that attempt to address the ill-posed problem by using a regularizer [48] or a priori knowledge of the sample, such as by assuming sparsity in its total variation [51, 52]. However, due to their inherent assumptions, these methods work well only on a limited range of samples, and the parameters defining the a priori knowledge need to be manually tuned to produce successful results. Fundamentally, they do not have the information in the spatial frequency regions where the OTF is zero, and the a priori knowledge attempts to fill in the missing gaps. We adopt a method proposed by Zhou et al. [31] which can capture all the spatial frequency information within the NA-limited bandpass, and faithfully recover the information via deconvolution without relying on a priori knowledge. A coded aperture designed by Zhou et al. at the pupil plane with a defocus aberration can generate a PSF whose OTF does not have zero values within its NA-limited bandpass. The particular coded aperture is generated by a genetic algorithm which searches for a binary mask pattern that maximizes its OTF's spatial frequency content's modulus. The optimum aperture's pattern is different depending on the amount of noise in the imaging condition. We choose the pattern as shown in Fig. 4 since it performs well across various noise levels [31].

The pupil function in our imaging scenario does not only consist of defocus, as the imaging lenses have severe aberrations. Therefore, our pupil function can have an unsymmetrical phase profile unlike the defocus aberration's symmetric bullseye phase profile. Thus, rotating the coded aperture can generate PSFs with different spatial frequency distribution beyond mere rotation. Therefore, in the second part of the data capturing procedure, we capture a series of images with a sequence of big masks, $A_n(u,v)$, consisting of 4 coded apertures and a standard circular aperture at the Fourier plane, as represented in Fig. 2 c). This ensure that we obtain all spatial frequency information within the NA-limited bandpass. The PSF associated with each $A_n(u,v)$ applied to $P_{t0}(u,v)$ is easily obtained by: $h_{n,t0}(\xi,\eta) = |\mathcal{F}^{-1}\{A_n(u,v)P_{t0}(u,v)\}(\xi,\eta)|^2$; and its OTF by: $H_{n,t0}(u,v) = \mathcal{F}\{h_{n,t0}(\xi,\eta)\}(u,v)$. With the measured full and coded aperture images, $\phi_{n,t0}(\xi,\eta)$, and the knowledge of the OTFs, we perform a combined deconvolution using iterative Tikhonov regularization, similar to Algorithm 1, to recover the object's intensity distribution, $o_{t0}(x,y)$, as described in Algorithm 3 and represented in Fig. 2 d).

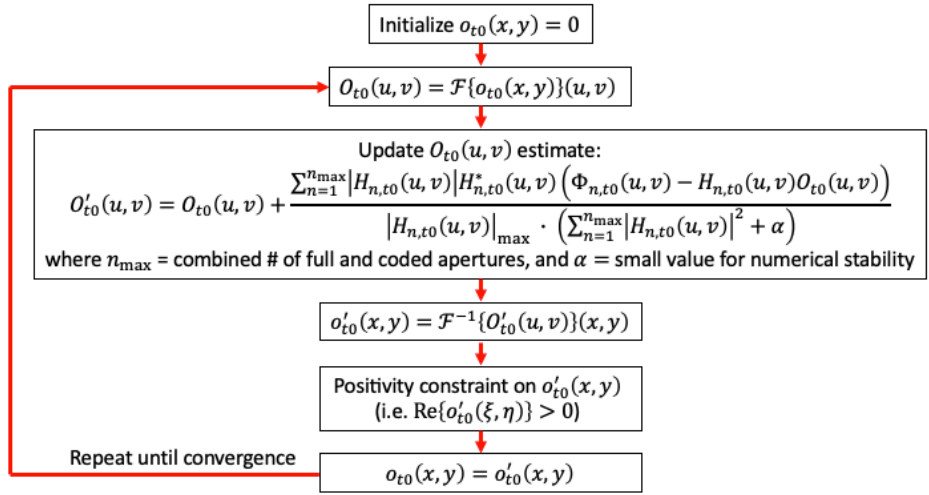

Algorithm 3. Iterative Tikhonov regularization for recovering the latent sample image, $o_{t0}(x,y)$, from aberrated images. Here, $\Phi_{n,t0}(u,v) = \mathcal{F}\{\phi_{n,t0}(\xi,\eta)\}(u,v)$.

A simulated deconvolution procedure with the coded aperture on a Siemens star pattern is shown in Fig. 4. The combined OTF of a circular aperture and the coded aperture at 4 rotation angles is able to eliminate the null regions found in the circular-aperture-only OTF and thus produce a superior deconvolution result. Experimentally, we observe that the signal-to-noise ratio (SNR) of at least 40 in the initial aberrated image produces a deconvolution result with minimal artifacts.

Section 2.1 - Demonstration of CACAO-FP on a crude lens

Experimental Setup

The CACAO-FP system's setup is simple, as shown in Fig. 5. It consists of a pair of 2-inch, f = 100 mm achromatic doublets (AC508-100-A from Thorlabs) to relay the surface of an imaging lens of interest to the surface of the ferroelectric liquid-crystal-on-silicon (LCOS) spatial light modulator (SLM) (SXGA-3DM-HB from 4DD). A polarized beam splitter (PBS) lies in front of the SLM to enable binary modulation of the SLM. A polarizer is placed after the PBS to further filter the polarized light to compensate for the PBS's low extinction ratio in reflection. A

f=200mm tube lens (TTL200-A from Thorlabs) Fourier transforms the modulated light and images it on a sCMOS camera (PCOedge 5.5 CL from PCO). The imaging system to be surveyed is placed in front of the CACAO-FP system at the first relay lens's focal length. The imaging system consists of a crude lens and a sample it is supposed to image. The crude lens in our experiment is a +6D trial lens (26 mm diameter) in an inexpensive trial lens set (TOWOO TW-104 TRIAL LENS SET). A sample consisting of an array of USAF targets is placed at less than the lens's focal length away to introduce more aberration into the system. The sample is flood-illuminated by a monochromatic LED light source (520 nm, model number) filtered with a 10nm-bandpass filter.

The relayed lens surface is modulated with various binary patterns by the SLM. The SLM displays a full aperture (5.5 mm diameter), a coded aperture rotated at 0°, 45°, 90°, and 135° with the maximum diameter matching the full aperture, and a series of limited apertures (1 mm diameter) shifted to different positions in a spiraling-out pattern within the full aperture dimension. The camera's exposure is triggered by the SLM for synchronization. Another trigger signal for enabling the camera to begin a capture sequence is provided by a data acquisition board (NI ELVIS II from National Instrument) which a user can control with MATLAB. Multiple images for each SLM aperture are captured and summed together to increase their signal-to-noise ratio (SNR). The full-aperture image had SNR=51, with other aperture-scanned images having SNR approximately proportional to the square root of their relative aperture area.

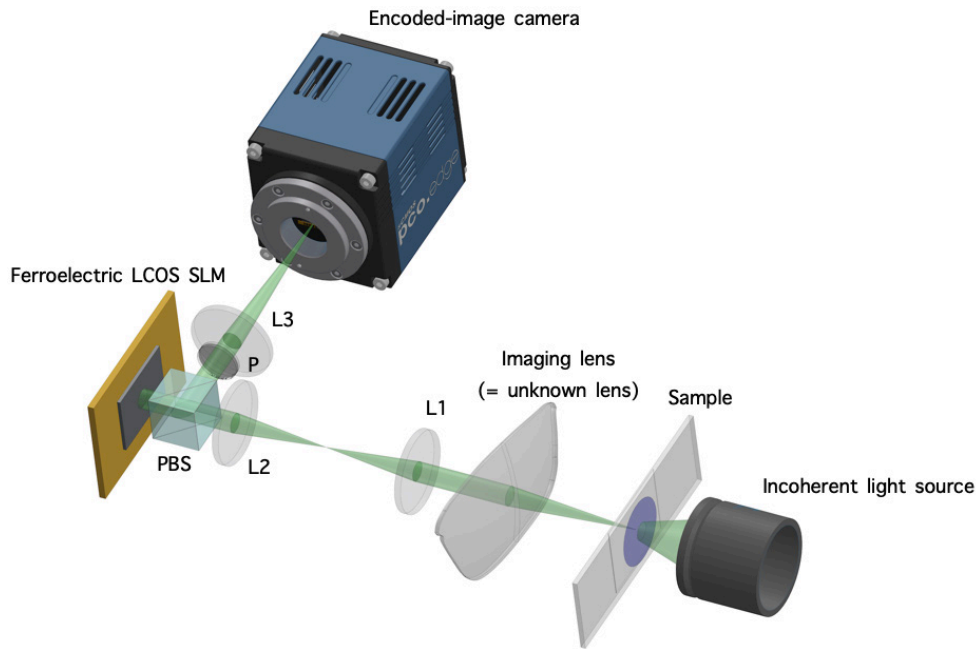

Figure 5. Experimental setup of imaging a sample with a crude lens (i.e. unknown lens). Sample is illuminated by a monochromatic LED (520 nm), and the lens' surface is imaged onto the SLM by a 1:1 lens relay. The part of light modulated by the SLM is reflected by the PBS and is further filtered by a polarizer to account for PBS's low extinction ratio in reflection (1:20). Pupil-modulated image of the sample is captured on the sCMOS camera.

The +6D lens is expected to have poor imaging performance that varies across its FOV since it is an inexpensive single element lens. We select 3 tiled regions, each containing a USAF target pattern, to demonstrate CACAO-FP's ability to address spatially variant aberration in its latent image recovery procedure. The expected resolution is between $\frac{520 \text{ nm}}{\text{NA}}$ (coherent) and $\frac{520 \text{ nm}}{2\text{NA}}$ (incoherent) periodicity, defined by the cut-off in the coherent and incoherent transfer functions, respectively. With the full aperture's diameter of d = 5.5 mm and the lens's focal length of f = 1/6 m = 167 mm, this corresponds to a range between Group 5 Element 1 and Group 5 Element 5. These elements are well-resolved across the three regions as shown in Fig. 6.

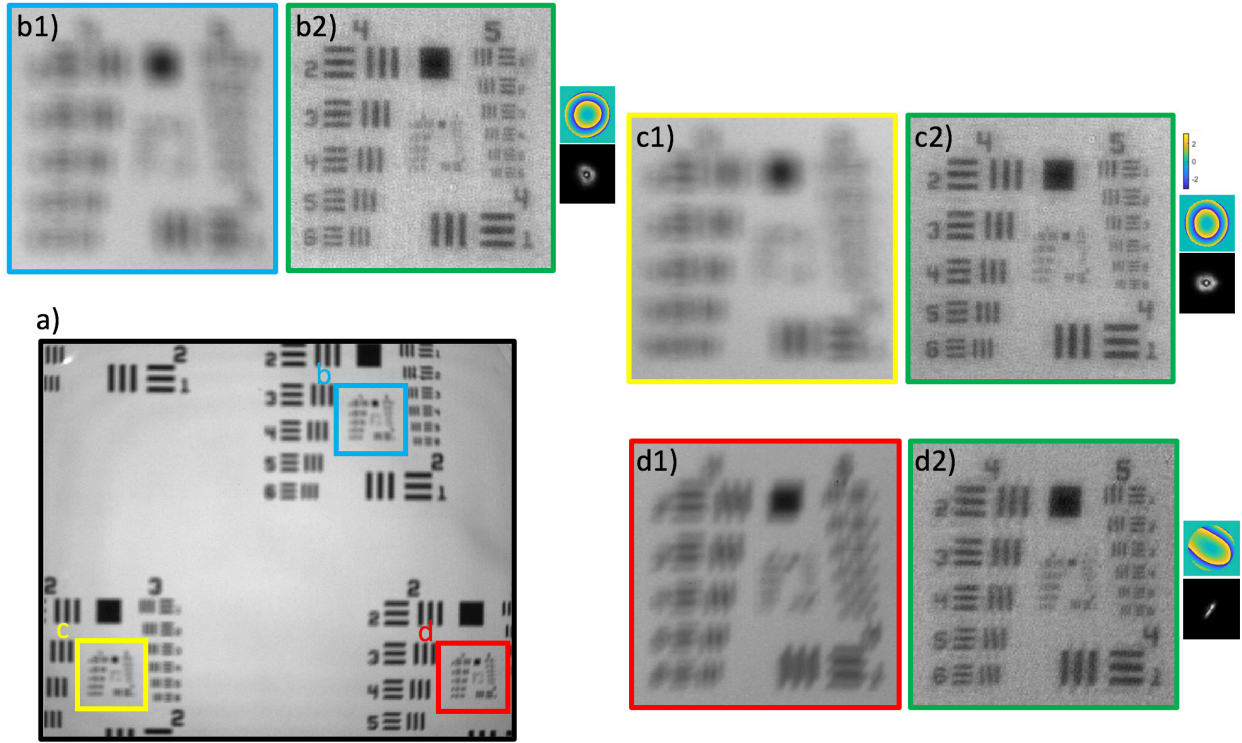

Figure 6. Spatially varying aberration compensation result on a grid of USAF target. a) the full FOV captured by our camera with the full circular aperture at 5.5 mm displayed on the SLM. Each small region denoted by b-d1) had a different aberration map as indicated by varying pupil function and PSFs. Spatially varying aberration is adequately compensated in post-processing as shown by the deconvolution results b-d2).

Section 2.2 - Demonstration of CACAO-FP on an eye phantom

Experimental Setup

We used an eye phantom from Ocular Instruments to simulate an in-vivo retinal imaging experiment. We embedded a cut-out USAF resolution target (2015a USAF from Ready Optics) on the phantom's retina, and filled the phantom's chamber with de-ionized water, as shown in Fig. 7. We made adjustments to our CACAO-FP system as shown in Fig. 8 to accommodate the different imaging scenario. First, it had to illuminate the retina in a reflection geometry via the same optical path as imaging. A polarized beam splitter (PBS) was used to provide illumination such that the specular reflection from the eye's cornea, which mostly maintains the s-polarization from the PBS, is filtered out of the imaging optical path. The scattered light from the retina is depolarized and can partially transmit through the PBS. The light source is a fiber-coupled laser diode (520 nm) which is made spatially incoherent by propagating through a 111-meter-long, 600-micron-core-diameter multimode fiber, following the method in Ref. [53]. The laser diode is triggered such that it is on only during camera exposure. We added a pupil camera that outputs the image of the eye's pupil with fiduciary marks for aligning the eye's pupil with our SLM. Finally, a motion-reference camera (MRC) that has the identical optical path as the encoded-image camera (EIC) aside from pupil-modulation by SLM is added to the system to account for an in-vivo eye's motion between image frames. The amount of light split between the MRC and EIC can be controlled by the PBS and a quarter-wave plate before the SLM.

In Fig. 9, the recovered images show severe spatially varying aberration of the eye phantom, but good deconvolution performance throughout the FOV, nonetheless. The full aperture in this scenario had 4.5-mm diameter, and its associated aberrated image had SNR of 126.

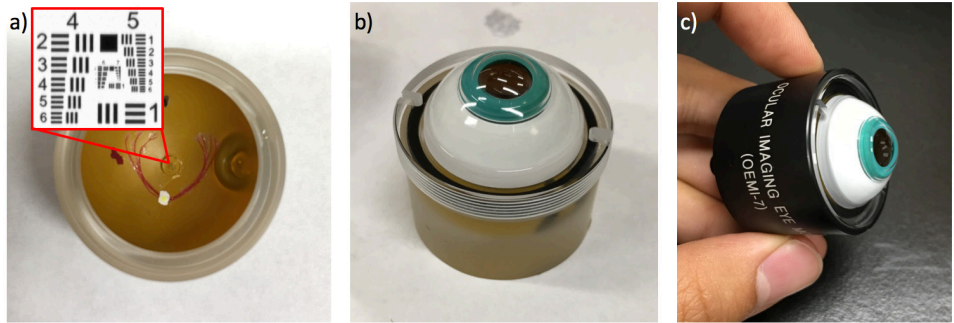

Figure 7. Eye phantom with a USAF target embedded on the retinal plane. a) A cut-out piece of glass of USAF target is attached on the retina of the eye phantom. The lid simulates the cornea and also houses a lens element behind it. b) The phantom is filled with water with no air bubbles in its optical path. c) The water-filled phantom is secured by screwing it in its case.

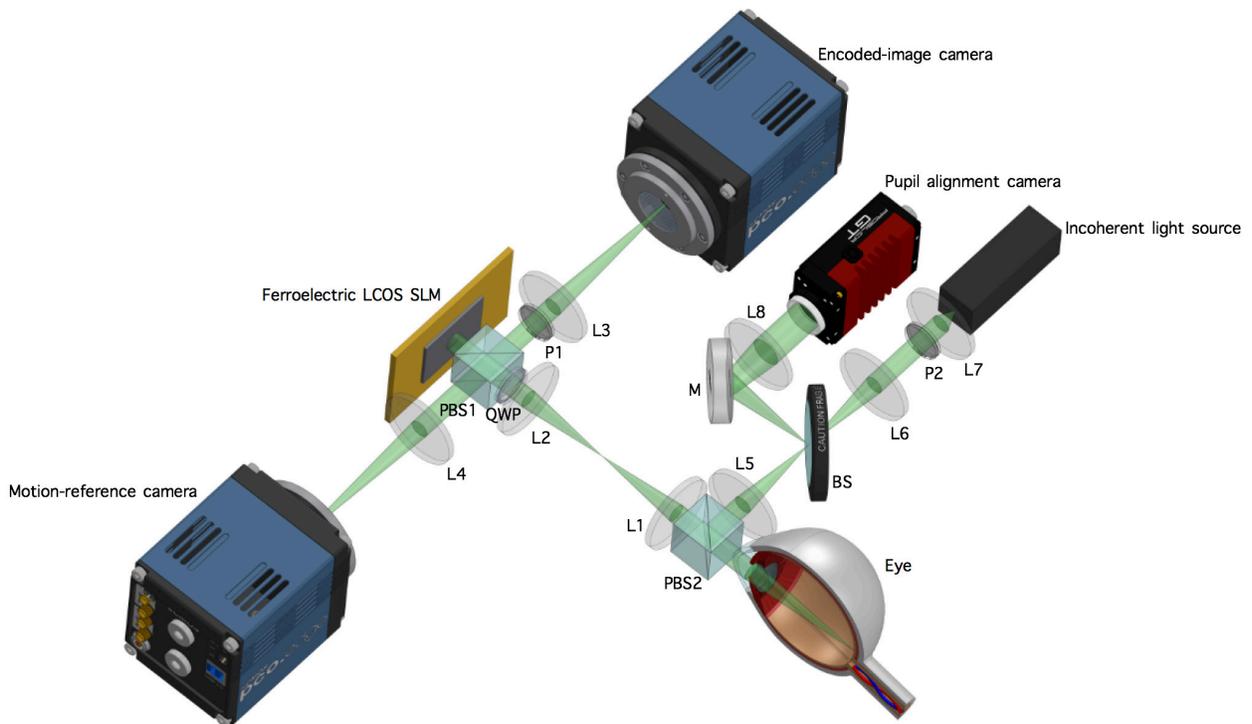

Figure 8. Experimental setup of imaging an eye phantom and an in-vivo eye. Illumination is provided by a fiber-coupled laser diode (520 nm), and the eye's pupil is imaged onto the SLM by a 1:1 lens relay. Pupil alignment camera provides fiduciary to the user for adequate alignment of the pupil on the SLM. PBS2 helps with removing corneal reflection. Motion-reference camera is synchronized with encoded-image camera to capture images not modulated by the SLM.

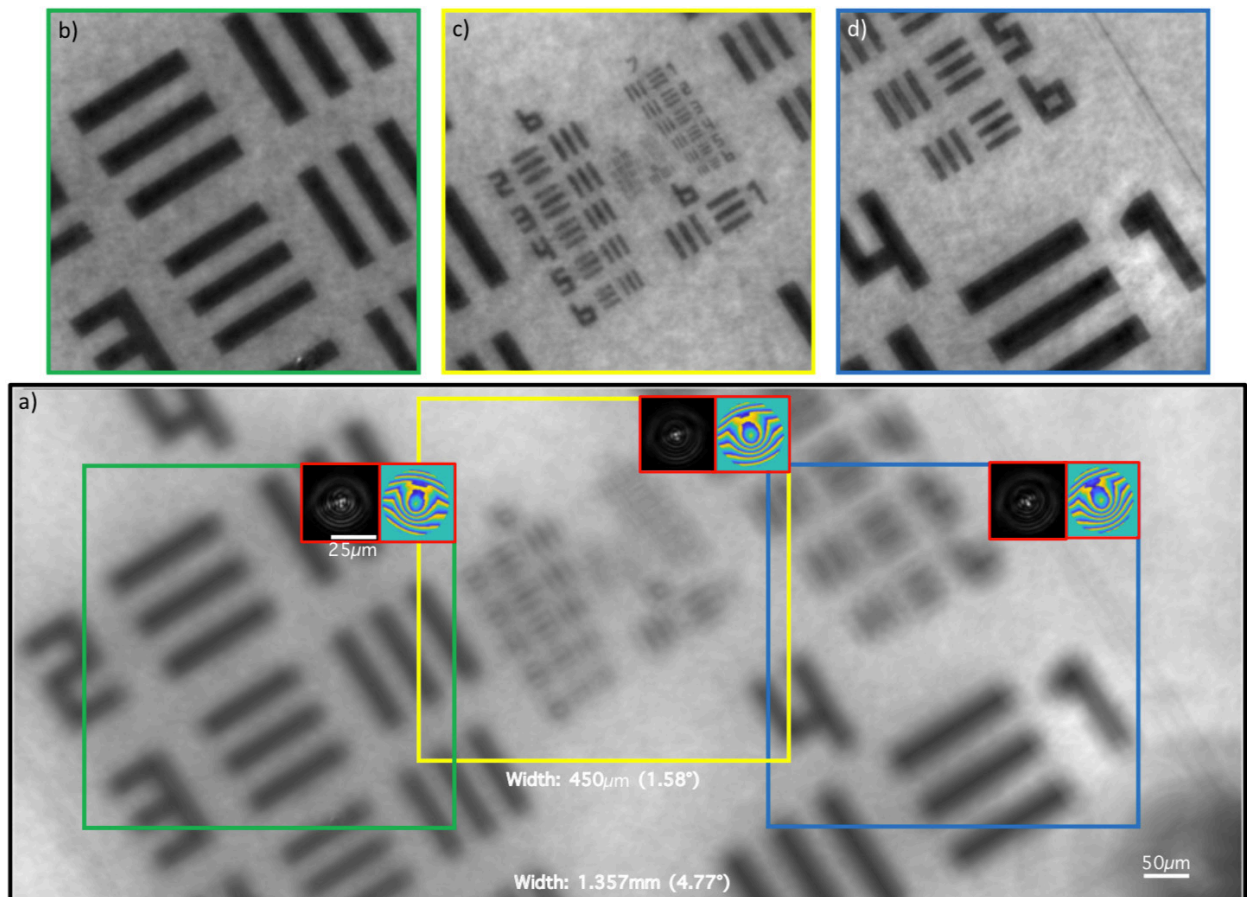

Figure 9. CACAO-FP result of imaging the USAF target in the eye phantom. a) Raw image averaged over 12 frames captured with the full circular aperture at 4.5 mm. The pupil function and PSF in each boxed region show the spatially varying aberration. b-d) Deconvolution results show sharp features of the USAF target. The uneven background is from the rough surface of the eye phantom's retina.

In this imaging scenario, the blur kernels of the limited-aperture images had a significant impact on the deconvolution result, as shown in Fig. 10. The aberration of the eye phantom was severe such that the retrieved blur kernels of the limited-aperture images had distinct shapes in addition to lateral shifts. We observe a much better deconvolution result with the reconstructed pupil that takes blur kernels' shapes into account compared to the one that does not. The latter is analogous to Shack-Hartmann wavefront sensing method, which only identifies the centroid of each blur kernel to estimate the aberration. Thus, this demonstrates the importance of the blur kernel estimation step in our algorithm and the distinct difference of our aberration reconstruction from other wavefront sensing methods.

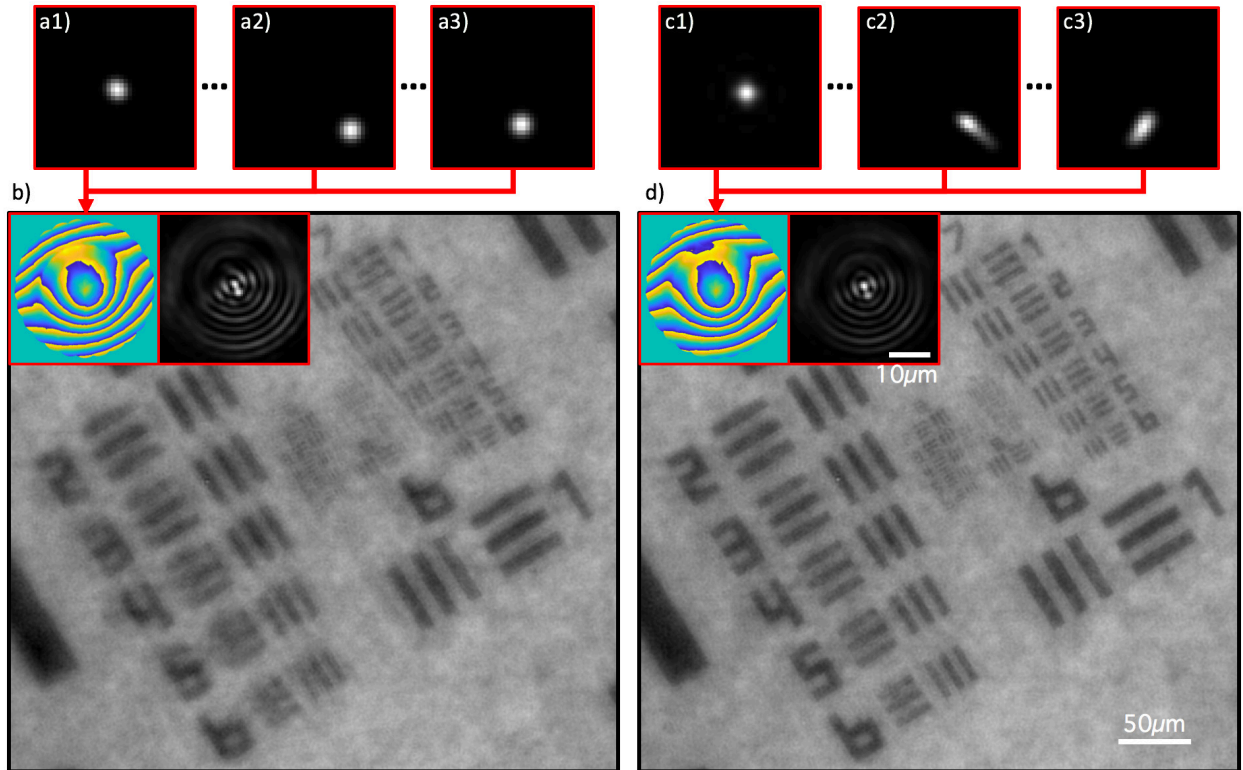

Figure 10. Showing the importance of masked pupil kernel shape determination for successful deconvolution. a1-3) limited PSFs determined only by considering their centroids. b) Recovered aberration and deconvolution result obtained with centroid-only limited PSFs. Some features of USAF are distorted. c1-3) limited PSFs determined with the blur estimation algorithm. d) Recovered aberration and deconvolution result obtained with the blur-estimated local PSFs. No distortions in the image are present, and more features of the USAF target are resolved.

Section 3 - Adapting CACAO-FP to an in-vivo experiment on the eye of a rhesus macaque

Experimental setup

The same setup as in Section 2.2 is used for the in-vivo experiment on a rhesus macaque's eye. The animal is anesthetized with 8-10mg/kg ketamine and 0.02mg/kg dexdomitor IM. 2 drops of tropicamide (0.5-1%) are placed on the eye to dilate the pupil. To keep the eye open for imaging, a sanitized speculum is be placed between the eyelids. A topical anesthetic (proparacaine 0.5%) is applied to the eye to prevent any irritation from the speculum placement. A rigid gas permeable lens will be placed on the eye to ensure that the cornea stays moist throughout imaging. The light intensity is kept below a level of ~ 50 mW/cm2 on the retina in accordance with ANSI recommended safe light dosage.

Due to the safety limitation on the illumination power, the increase of captured images' SNR is achieved by acquiring many redundant frames for summation. This meant that a long sequence of images (30 seconds) had to be captured, and these images had to be registered for motion prior to CACAO-FP (S.Vid. 1) since the eye had residual motion even under anesthesia. Due to a long averaging window, aberration of high temporal frequency is washed out, but we still expect to be able to resolve the photoreceptors albeit at a lower contrast [54]. Motion registration included both translation and rotation, and these operations needed to be done such that they do not apply any spatial filter that may alter the images' spatial frequency spectra. Rotation is performed with fast discrete sinc-interpolation [55] which is a series of Fourier transform operations that can be accelerated by GPU programming. The frames from the motion-reference camera are used for the registration process (S.Vid. 1), and the extracted registration parameters are applied to the images of the encoded-image camera.

The deconvolution result is shown in Fig. 11. The input full-aperture image had SNR=109, with little variation across its FOV. Photoreceptors are much better resolved after aberration removal. We expect the entire visible region to have an even spread of photoreceptors, but we observe well-resolved photoreceptors mostly in the brighter regions. This may be due to the lower SNR in the darker regions leading to poorer deconvolution result. Despite the high SNR compared to our eye phantom experiment, the deconvolution result is less optimal due to the loss of detail from less-than-perfect motion registration and washing-out of high-temporal-frequency aberration during averaging over multiple frames.

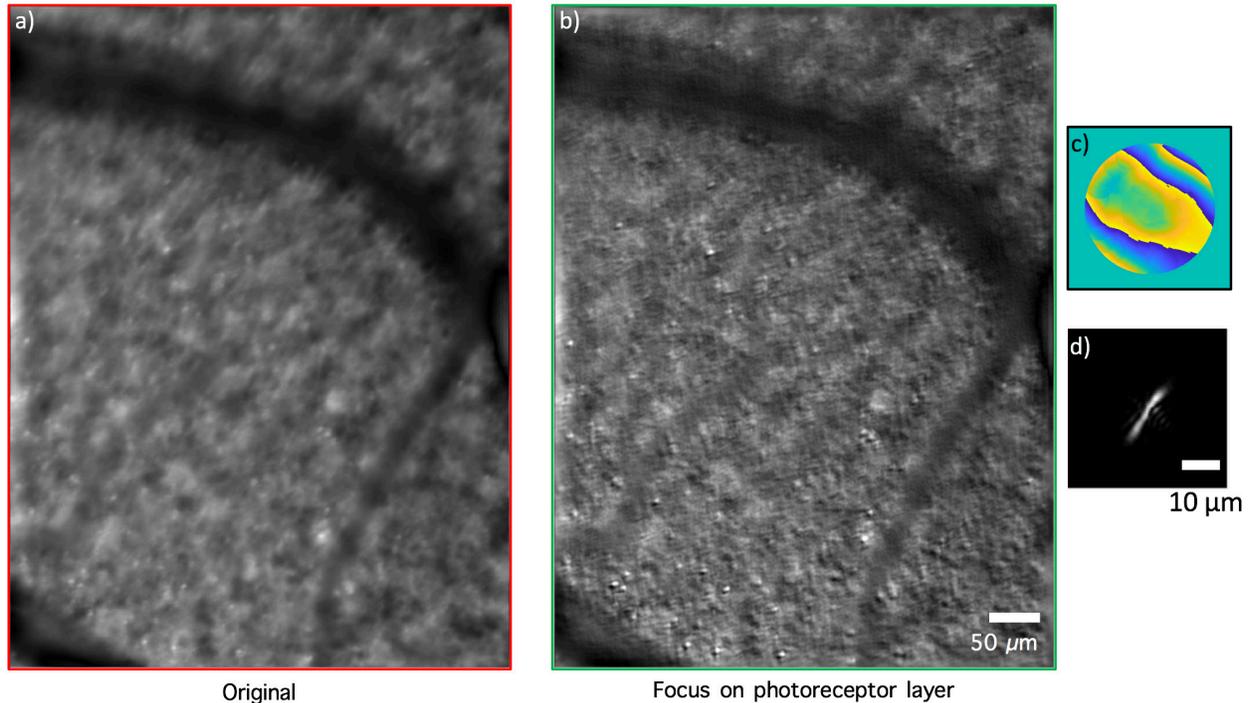

Figure 11. CACAO-FP result from imaging an in-vivo eye of a rhesus macaque. a) Raw image averaged over 213 frames captured with 4.5mm full circular aperture. b) Deconvolution result using the c) pupil function reconstructed by CACAO-FP procedure. d) PSF associated with the pupil function.

Discussion

We developed a novel method to computationally correct for aberration of a crude lens without spatial coherence requirement. Our demonstration of CACAO-FP on sub-optimal lenses in benchtop and in-vivo experiments shows its viability in a broad range of imaging scenarios. Its simple hardware setup is also a key advantage over other aberration-correction methods that may allow for wider adoption.

More severe aberration can be addressed readily by shrinking the scanned aperture size on the SLM so that the aberration within the windowed pupil function does not become severe. This comes at the expense of the acquisition speed as more images need to be captured to cover the same pupil diameter.

If the masks on the pupil are shrunk smaller with no overlap, this pupil masking process becomes a Shack-Hartmann (SH) sensing method. This illustrates the key advantages of our scheme over a SH sensor: using bigger masks allows for a fewer number of image acquisitions and increases the images' SNR. A bigger mask of an aberrated pupil no longer encodes for a simple shifted spot in the spatial domain as would be the case for a SH sensor, but rather a blur kernel as shown in Fig. 3. Therefore, reconstructing the blur kernels of the limited aperture images is critical for CACAO-FP's performance, as is demonstrated in section 2.2.

The retinal image recovered with CACAO-FP in section 3 was not as well resolved as images one would expect from a typical adaptive optics retinal imager due to a couple of reasons. First, deconvolution is sensitive to noise and the increase in the input image SNR could only be achieved by averaging multiple frames that resulted in increased acquisition time. Despite the animal being under general anesthesia, its eye continued to drift during the acquisition process and placed a limit on the length of acquisition before the region of interest on the retina shifts out of the field of view. Imaging a human subject would be less susceptible to this issue as an operator can instruct the subject to focus on a target. Using a different wavelength invisible to the eye may also reduce the motion if the movement is due to glare aversion response. Second, the specular reflection from the retina corrupted the captured images. The flood illumination provided on the retina through the pupil did not have sufficient angular coverage to even out the specular reflection such that some images captured with a small aperture contained specular reflection while others did not. A flood illumination with wider divergence angle can mitigate this problem.

CACAO-FP can also be adapted to address chromatic aberration in a system by accounting for the aberration under each color separately and combining the different color channel images digitally. Geometrical distortions may have to be applied due to the change in the lens's focal length for different wavelengths.

Funding

National Institute of Health (NIH) Agency Award: R21 EY026228A.

Acknowledgements

We thank Amir Hariri for assisting with the experiment, Soo-Young Kim for generously sharing a fixed mouse retina sample, and Mooseok Jang, Haowen Ruan, Edward Haojiang Zhou, Joshua Brake, Michelle Cua, and Hangwen Lu for helpful discussions.

Appendix

Appendix 1:

We consider a point on the unknown sample, $s(x,y)$, and how it propagates to the camera plane to be imaged. On the sample plane, a point source at $(x_0, y_0)$ may have an amplitude and phase $C$, and it can be described by:

$$U_0(x, y; x_0, y_0) = C\delta(x - x_0, y - y_0)$$

S.Eq. 1

We then use Fresnel propagation:

$$U_1(u, v; x_0, y_0) = \frac{e^{j\frac{\pi}{\lambda f_0}(u^2+v^2)}}{j\lambda f_0} \int_{-\infty}^{\infty} C\delta(x - x_0, y - y_0) e^{j\frac{\pi}{\lambda f_0}(x^2+y^2)} e^{-j\frac{2\pi}{\lambda f_0}(xu+yv)} dx\, dy$$

$$= C\frac{e^{j\frac{\pi}{\lambda f_0}(u^2+v^2)}}{j\lambda f_0} e^{j\frac{\pi}{\lambda f_0}(x_0^2+y_0^2)} e^{-j\frac{2\pi}{\lambda f_0}(x_0 u + y_0 v)}$$

S.Eq. 2

and apply the phase delay associated with an idealized thin lens [45] having an estimated focal length $f_0$ for the unknown lens, $e^{-j\frac{\pi}{\lambda f_0}(u^2+v^2)}$, and any discrepancy from the ideal is incorporated into the pupil function, $P(u, v; x_0, y_0) = P_{t0}(u, v)$ where $t0$ is the isoplanatic patch around $(x_0, y_0)$:

$$U_2(u, v; x_0, y_0) = \frac{C}{j\lambda f_0} e^{j\frac{\pi}{\lambda f_0}(x_0^2+y_0^2)} e^{-j\frac{2\pi}{\lambda f_0}(x_0 u + y_0 v)} P_{t0}(u, v) = C_2(x_0, y_0) P_{t0}(u, v) e^{-j\frac{2\pi}{\lambda f_0}(x_0 u + y_0 v)}$$

S.Eq. 3

where we set $C_2(x_0, y_0) = \frac{C}{j\lambda f_0} e^{j\frac{\pi}{\lambda f_0}(x_0^2+y_0^2)}$. S.Eq. 3 is Eq. 2.

$U_2(u, v; x_0, y_0)$ is relayed to the Fourier plane in Fig. 1 without any additional phase term by the 4f system

formed by L_s1 and L_s2 [45]. The relayed field may be magnified by the factor $\frac{f_2}{f_1}$, but here we assume no magnification. We apply a mask, $M(u,v)$, to the field:

$$U_2'(u,v;x_0,y_0) = M(u,v)C_2(x_0,y_0)P_{t0}(u,v)e^{-j\frac{2\pi}{\lambda f_0}(x_0 u + y_0 v)}$$

S.Eq. 4

and propagate it by distance $d$ using angular spectrum to the surface of L_s3:

$$U_3(s,t;x_0,y_0) = \mathcal{F}^{-1}\left\{\mathcal{F}\{U_2'(u,v;x_0,y_0)\}e^{-j\frac{2\pi d}{\lambda}\sqrt{1-(\lambda f_x)^2 - (\lambda f_y)^2}}\right\}(s,t)$$

S.Eq. 5

where $(s,t)$ are the coordinates on the L_s3's plane. We then apply the phase delay associated with L_s3, $e^{-j\frac{\pi}{\lambda f_3}(s^2+t^2)}$, and propagate the field by $f_3$ to the camera plane:

$$U_4(\xi,\eta;x_0,y_0) = \frac{e^{j\frac{\pi}{\lambda f_3}(\xi^2+\eta^2)}}{j\lambda f_3} \int_{-\infty}^{\infty} e^{-j\frac{\pi}{\lambda f_3}(s^2+t^2)} U_3(s,t;x_0,y_0) e^{j\frac{\pi}{\lambda f_3}(s^2+t^2)} e^{-j\frac{2\pi}{\lambda f_3}(s\xi+t\eta)} ds\,dt$$

$$= \frac{e^{j\frac{\pi}{\lambda f_3}(\xi^2+\eta^2)}}{j\lambda f_3} \int_{-\infty}^{\infty} \mathcal{F}^{-1}\left\{\mathcal{F}\{U_2'(u,v;x_0,y_0)\}e^{-j\frac{2\pi d}{\lambda}\sqrt{1-(\lambda f_x)^2 - (\lambda f_y)^2}}\right\}(s,t)\, e^{-j\frac{2\pi}{\lambda f_3}(s\xi+t\eta)} ds\,dt$$

S.Eq. 6

Setting $(\xi',\eta') = \frac{1}{\lambda f_3}(\xi,\eta)$ and $C_4(\xi',\eta') = \frac{e^{j\pi\lambda f_3(\xi'^2+\eta'^2)}}{j\lambda f_3}$:

$$U_4'(\xi',\eta';x_0,y_0) = C_4(\xi',\eta')\mathcal{F}\left\{\mathcal{F}^{-1}\left\{\mathcal{F}\{U_2'(u,v;x_0,y_0)\}e^{-j\frac{2\pi d}{\lambda}\sqrt{1-(\lambda f_x)^2 - (\lambda f_y)^2}}\right\}\right\}(\xi',\eta')$$

$$= C_4(\xi',\eta')\mathcal{F}\left\{U_2'(u,v;x_0,y_0) * \mathcal{F}^{-1}\left\{e^{-j\frac{2\pi d}{\lambda}\sqrt{1-(\lambda f_x)^2 - (\lambda f_y)^2}}\right\}\right\}(\xi',\eta')$$

$$= C_4(\xi',\eta')\mathcal{F}\{U_2'(u,v;x_0,y_0)\}(\xi',\eta')e^{-j\frac{2\pi d}{\lambda}\sqrt{1-(\lambda\xi')^2 - (\lambda\eta')^2}}$$

$$= C_4(\xi',\eta')e^{-j\frac{2\pi d}{\lambda}\sqrt{1-(\lambda\xi')^2-(\lambda\eta')^2}}\mathcal{F}\left\{M(u,v)C_2(x_0,y_0)P_{t0}(u,v)e^{-j\frac{2\pi}{\lambda f_0}(x_0 u+y_0 v)}\right\}(\xi',\eta')$$

S.Eq. 7

Setting $C_5(\xi',\eta';x_0,y_0) = C_4(\xi',\eta')e^{-j\frac{2\pi d}{\lambda}\sqrt{1-(\lambda\xi')^2-(\lambda\eta')^2}}C_2(x_0,y_0)$:

$$U_4'(\xi',\eta';x_0,y_0) = C_5(\xi',\eta';x_0,y_0)\left[\mathcal{F}\{M(u,v)P_{t0}(u,v)\}(\xi',\eta') * \delta\left(\xi'+\frac{x_0}{\lambda f_0},\eta'+\frac{y_0}{\lambda f_0}\right)\right]$$

$$= C_5(\xi',\eta';x_0,y_0)\mathcal{F}\{M(u,v)P_{t0}(u,v)\}\left(\xi'+\frac{x_0}{\lambda f_0},\eta'+\frac{y_0}{\lambda f_0}\right)$$

S.Eq. 8

This is the complex field incident on the camera from the point source located at $(x_0,y_0)$. It is the PSF of the system, and we observe that it simply shifts laterally for different $(x,y)$ coordinates. Therefore, the image on the camera sensor can be calculated by a convolution between $U_4'(\xi',\eta';x_0,y_0)$ and the sample field within the isoplanatic patch, $s_{t0}(x,y)$. However, the phase term in $C_5(\xi',\eta';x_0,y_0)$ can have a significant impact on the captured images. In our incoherent imaging scenario, the phase relationship between the points on the sample plane during the capturing process is irrelevant. So, we can define an intensity PSF, $h_{t0}(\xi,\eta) = |\mathcal{F}\{M(u,v)P_{t0}(u,v)\}(\xi,\eta)|^2$, to describe the intensity of the $U_4'(\xi',\eta';x_0,y_0)$ captured by the camera:

$$|U_4'(\xi',\eta';x_0,y_0)|^2 = |C_5(\xi',\eta';x_0,y_0)|^2 h_{t0}\left(\xi'+\frac{x_0}{\lambda f_0},\eta'+\frac{y_0}{\lambda f_0}\right) = \frac{C}{\lambda^2 f_0 f_3}h_{t0}\left(\xi'+\frac{x_0}{\lambda f_0},\eta'+\frac{y_0}{\lambda f_0}\right)$$

S.Eq. 9

The complicated phase fluctuations embedded in $C_5(\xi', \eta'; x_0, y_0)$ become no longer relevant. Dropping the constants and neglecting coordinate scaling, the image of the unknown sample captured by the camera becomes a convolution of $h_{t0}(\xi, \eta)$ with the sample distribution, $s_{t0}(x, y)$, as described by Eq. 5.

Appendix 2

In processing the data from the in-vivo experiment, motion-reference camera images are first registered for rotation and translation, as shown in S.Vid. 1, and these registration values are taken into account when summing multiple frames captured with the same SLM aperture pattern.
An example of a single frame of full aperture image is shown in S.Fig. 1, and the same aperture image after summing 213 frames.

Due to the photon-starved imaging condition, it is imperative to account for the detector noise in the captured images. We use two-point radiometric calibration to account for the fixed pattern noise and inhomogeneous sensitivity of our imaging sensor [56]:

$$I' = \frac{I - B}{R - B}$$

S.Eq. 11

where $I'$ is the desired calibrated image, $I$ is the input image, $B$ is the dark image captured with the sensor blocked from light, and $R$ is a reference image captured with the sensor capturing an image of an opal diffuser.

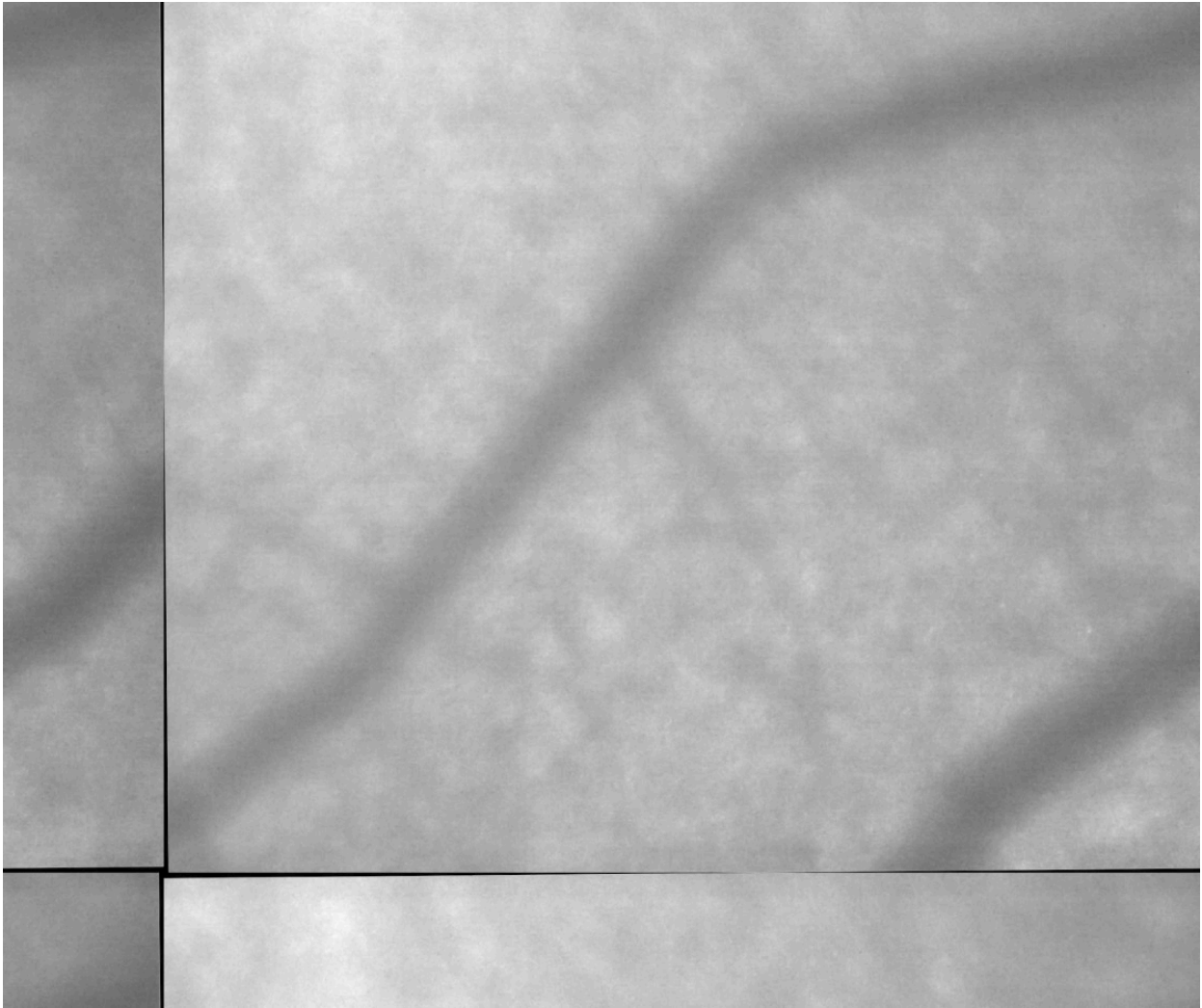
Supplementary Video 1. Registering image frames for rotation and translation.

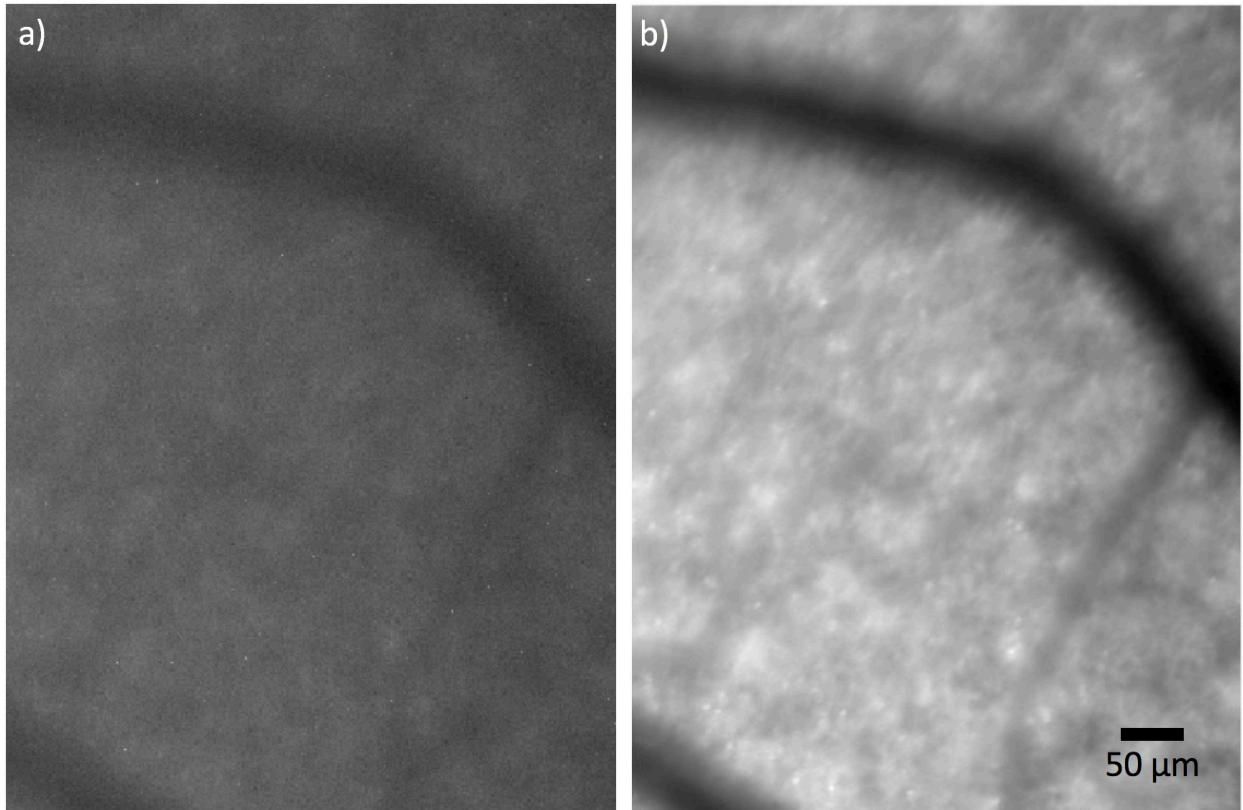

Supplementary Figure 1. Raw full aperture image. a) 1 frame and b) a sum of 213 frames.

References:

[1] A. W. Lohmann, R. G. Dorsch, D. Mendlovic, Z. Zalevsky, and C. Ferreira, "Space-bandwidth product of optical signals and systems," J. Opt. Soc. Am. A 13(3), 470-473 (1996).
[2] G. McConnell, J. Trägårdh, R. Amor, J. Dempster, E. Reid, and W. B. Amos, "A novel optical microscope for imaging large embryos and tissue volumes with sub-cellular resolution throughout," eLife 5, e18659 (2016).
[3] P. Godara, A. Dubis, A. Roorda, J. Duncan, and J. Carroll, "Adaptive Optics Retinal Imaging: Emerging Clinical Applications," Optometry & Vision Science 87(12), 930 (2010).
[4] D. Williams, "Imaging single cells in the living retina," Vision Research 51(13), 1379–1396 (2011).
[5] M. J. Booth, "Adaptive optical microscopy: the ongoing quest for a perfect image," Light: Science & Applications (2014) 3, e165 (2014).
[6] S. Marcos, J. S. Werner, S. A. Burns, W. H. Merigan, P. Artal, D. A. Atchison, K. M. Hampson, R. Legras, L. Lundstrom, G. Yoon, J. Carroll, S. S. Choi, N. Doble, A. M. Dubis, A. Dubra, A. Elsner, R. Jonnal, D. T. Miller, M. Paques, H. E. Smithson, L. K. Young, Y. Zhang, M. Campbell, J. Hunter, A. Metha, G. Palczewska, J. Schallek, and L. C. Sincich, "Vision science and adaptive optics, the state of the field," Vis. Res. 132, 3–33 (2017).
[7] G. Zheng, R. Horstmeyer, and C. Yang, "Wide-field, high-resolution Fourier ptychographic microscopy," Nat Photonics 7(9), 739–745 (2013).
[8] R. Horstmeyer, X. Ou, J. Chung, G. Zheng, and C. Yang, "Overlapped Fourier coding for optical aberration removal," Opt. Express 22(20), 24062–24080 (2014).
[9] X. Ou, G. Zheng, and C. Yang, "Embedded pupil function recovery for Fourier ptychographic microscopy," Opt. Express 22(5), (2014).
[10] X. Ou, R. Horstmeyer, C. Yang, and Guoan Zheng, "Quantitative phase imaging via Fourier ptychographic microscopy," Optics Letters 38(22), 4845-4848, (2013).
[11] Z. Bian, S. Dong, and G. Zheng, "Adaptive system correction for robust Fourier ptychographic imaging," Opt. Express 21(26), 32400-32410 (2013).

[12] L. Bian, J. Suo, J. Chung, X. Ou, C. Yang, F. Chen, and Q. Dai, "Fourier ptychographic reconstruction using Poisson maximum likelihood and truncated Wirtinger gradient," Sci. Rep. 6, 27384 (2016).
[13] L. Bian, J. Suo, G. Zheng, K. Guo, F. Chen, and Q. Dai, "Fourier ptychographic reconstruction using Wirtinger flow optimization," Opt. Express 23(4), 4856-4866 (2015).
[14] L. H. Yeh, J. Dong, J. Zhong, L. Tian, and M. Chen, "Experimental robustness of Fourier ptychography phase retrieval algorithms," Opt. Express 23(26), 33214-33240 (2015).
[15] L. Bian, J. Suo, G. Situ, G. Zheng, F. Chen, and Q. Dai, "Content adaptive illumination for Fourier ptychography," Optics Letters 39(23), 6648-6651 (2014).
[16] J. Sun, Q. Chen, Y. Zhang, and C. Zuo, "Efficient positional misalignment correction method for Fourier ptychographic microscopy," Biomed. Opt. Express 7(4), 1336-1350 (2016).
[17] Y. Zhang, W. Jiang, L. Tian, L. Waller, and Q. Dai, "Self-learning based Fourier ptychographic microscopy," Opt. Express 23(14), 18471-18486 (2015).
[18] P. Li, D. J. Batey, T. B. Edo, and J. M. Rodenburg, "Separation of three-dimensional scattering effects in tilt-series Fourier ptychography," Ultramicroscopy 158, 1-7 (2015).
[19] R. Horstmeyer, J. Chung, X. Ou, G. Zheng, and C. Yang, "Diffraction tomography with Fourier ptychography," Optica 3(8), 827-835 (2016).
[20] T Kamal, L Yang, and L.-W. express, "In situ retrieval and correction of aberrations in moldless lenses using Fourier ptychography," Optics Express (2018).
[21] X. Ou, R. Horstmeyer, G. Zheng, and C. Yang, "High numerical aperture Fourier ptychography: principle, implementation and characterization," Optics Express 23(3), 3472–3491 (2015).
[22] J Chung, J Kim, X Ou, R. Horstmeyer, and C. Yang, "Wide field-of-view fluorescence image deconvolution with aberration-estimation from Fourier ptychography," Biomedical Optics Express (2016).
[23] L. Tian, X. Li, K. Ramchandran, and L. Waller, "Multiplexed coded illumination for Fourier Ptychography with an LED array microscope," Biomedical optics express 5(7), 2376–2389 (2014).
[24] J. Chung, H. Lu, X. Ou, H. Zhou, and C. Yang, "Wide-field Fourier ptychographic microscopy using laser illumination source," Biomedical optics express 7(11), 4787–4802 (2016).
[25] L. Tian and L. Waller, "3D intensity and phase imaging from light field measurements in an LED array microscope," Optica 2(2), 104–111 (2015).
[26] L. Tian, Z. Liu, L.-H. Yeh, M. Chen, J. Zhong, and L. Waller, "Computational illumination for high-speed in vitro Fourier ptychographic microscopy," Optica 2(10), 904-911 (2015).
[27] A. Williams, J. Chung, X. Ou, G. Zheng, S. Rawal, Z. Ao, R. Datar, C. Yang, and R. Cote, "Fourier ptychographic microscopy for filtration-based circulating tumor cell enumeration and analysis," J. Biomed. Opt. 19(6), 066007 (2014).
[28] S. Dong, K. Guo, P. Nanda, R. Shiradkar, and G. Zheng, "FPscope: a field-portable high-resolution microscope using a cellphone lens," Optics Express 5(10), 3305-3310 (2014).
[29] J. Sun, C. Zuo, L. Zhang, and Q. Chen, "Resolution-enhanced Fourier ptychographic microscopy based on high-numerical-aperture illuminations," Sci. Rep. 7, 1187 (2017).
[30] C. Kuang, Y. Ma, R. Zhou, J. Lee, G. Barbastathis, R. R. Dasari, Z. Yaqoob, and P. T. C. So, "Digital micromirror device-based laser-illumination Fourier ptychographic microscopy," Opt. Express 23(21), 26999-27010 (2015).
[31] C. Zhou and S. Nayar, "What are good apertures for defocus deblurring?" 2009 IEEE International Conference on Computational Photography.
[32] Fienup and Miller, "Aberration correction by maximizing generalized sharpness metrics," J Opt Soc Am 20(4), 609 (2003).
[33] D. Hillmann, H. Spahr, C. Hain, H. Sudkamp, G. Franke, C. Pfäffle, C. Winter, and G. Hüttmann, "Aberration-free volumetric high-speed imaging of in vivo retina," Scientific Reports 6(1), 35209 (2016).
[34] F. Soulez, L. Denis, Y. Tourneur, and É. Thiébaut, "Blind deconvolution of 3D data in wide field fluorescence microscopy," 2012 9th IEEE International Symposium on Biomedical Imaging (ISBI).
[35] E. Thiébaut and J.-M. Conan, "Strict a priori constraints for maximum-likelihood blind deconvolution," J. Opt. Soc. Am. A 12(3), 485-492 (1995).
[36] S. Adie, B. Graf, A. Ahmad, S. Carney, and S. Boppart, "Computational adaptive optics for broadband optical interferometric tomography of biological tissue," Proceedings of the National Academy of Sciences 109(19), 7175–7180 (2012).
[37] N. Shemonski, F. South, Y.-Z. Liu, S. Adie, S. Carney, and S. Boppart, "Computational high-resolution optical imaging of the living human retina," Nature Photonics 9(7), 440–443 (2015).

[38] D. Kundur and D. Hatzinakos, "Blind image deconvolution," IEEE Signal Processing Magazine 13(2), 43-64 (1996).
[39] A. Kumar, W. Drexler, and R. A. Leitgeb, "Subaperture correlation based digital adaptive optics for full field optical coherence tomography," Opt. Express 21(9), 10850–10866 (2013).
[40] A. Kumar, D. Fechtig, L. Wurster, L. Ginner, M. Salas, M. Pircher, and R. Leitgeb, "Noniterative digital aberration correction for cellular resolution retinal optical coherence tomography in vivo," Optica 4(8), 924–931 (2017).
[41] L. Ginner, T. Schmoll, A. Kumar, M. Salas, N. Pricoupenko, L. Wurster, and R. Leitgeb, "Holographic line field en-face OCT with digital adaptive optics in the retina in vivo," Biomedical Optics Express 9(2), 472 (2018).
[42] G. Gunjala, S. Sherwin, A. Shanker, and L. Waller, "Aberration recovery by imaging a weak diffuser," Optics Express 26(16), 21054-21068 (2018)
[43] G. Zheng, X. Ou, R. Horstmeyer, and C. Yang, "Characterization of spatially varying aberrations for wide field-of-view microscopy," Opt. Express 21(13), 15131-15143 (2013).
[44] B. K. Gunturk and X. Li, Image Restoration: Fundamentals and Advances (CRC Press, 2012), Vol. 7.
[45] J. Goodman, Introduction to Fourier Optics (McGraw-Hill, 2008).
[46] L. Yuan, J. Sun, L. Quan, and H.-Y. Shum, "Image deblurring with blurred/noisy image pairs," ACM Transactions on Graphics (TOG) 26(3), 1 (2007).
[47] S. H. Lim and D. A. Silverstein, Method for deblurring an image. US Patent Application, Pub. No. US2006/0187308 A1, Aug 24, 2006.
[48] A. Neumaier, "Solving ill-conditioned and singular linear systems: A tutorial on regularization," SIAM review (1998).
[49] Rodenburg and Faulkner, "A phase retrieval algorithm for shifting illumination," Applied Physics Letters 85(20), 4795–4797 (2004).
[50] J. Sun, Q. Chen, Y. Zhang, and C. Zuo, "Sampling criteria for Fourier ptychographic microscopy in object space and frequency space," Optics Express 24(14), 15765-15781 (2016).
[51] J. M. Bioucas-Dias, M. A. T. Figueiredo, and J. P. Oliveira, "Total variation-based image deconvolution: a majorization-minimization approach," 2006 IEEE International Conference on Acoustics Speech and Signal Processing Proceedings.
[52] A. Levin, R. Fergus, F. Durand, and W. T. Freeman, "Image and Depth from a Conventional Camera with a Coded Aperture," CM Transactions on Graphics (TOG) 26(3), 70 (2007).
[53] J. Rha, R. S. Jonnal, K. E. Thorn, J. Qu, Y. Zhang, and D. T. Miller, "Adaptive optics flood-illumination camera for high speed retinal imaging," Optics Express 14(10), 4552-4569 (2006).
[54] H. Hofer, L. Chen, G. Y. Yoon, B. Singer, Y. Yamauchi, and D. R. Williams, "Improvement in retinal image quality with dynamic correction of the eye's aberrations," Opt. Express 8(11), 631-643 (2001).
[55] L. Yaroslavsky, Theoretical Foundations of Digital Imaging Using MATLAB (CRC Press, 2013).
[56] B. Jähne, Digital Image Processing (Springer, 2005).